# Hierarchical spectral inhomogeneity in photoluminescence of a twisted $MoSe_2/WSe_2$ heterobilayer moiré superlattice revealed by hyperspectral mapping

*Nurul Fariha Ahmad*[1], *Yuto Urano*[1], *Kenji Watanabe*[3], *Takashi Taniguchi*[1], *Daichi Kozawa*[1], *and Ryo Kitaura*[1,2,*]

[1] *Research Center for Materials Nanoarchitectonics, National Institute for Materials Science, 1-1 Namiki, Tsukuba-city, Ibaraki 305-0044, Japan*

[2] *Graduate School of Chemical Sciences and Engineering, Hokkaido University, 5, Kita 8 Nishi, Kita-ku, Sapporo-city, Hokkaido 060-8628, Japan*

[3] *Research Center for Electronic and Optical Materials, National Institute for Materials Science, 1-1 Namiki, Tsukuba-city, Ibaraki 305-0044, Japan*

* Correspondence to KITAURA.Ryo@nims.go.jp

**Abstract**

Low-temperature photoluminescence from $MoSe_2/WSe_2$ moiré superlattice often consists of a broad interlayer emission background with dense, narrow peaks, making microscopic line-by-line assignment difficult. Here, we use hyperspectral photoluminescence mapping and peak-decomposition-free spectral analyses to determine how this spectral complexity is organized in space. A 20 × 20 map acquired with a 400 nm pitch reveals three dominant spectral families that form contiguous real-space domains. Feature-wise spatial correlation analysis and whole-spectrum similarity yield a characteristic micron-scale length of 1.27–2.05 μm, all exceeding the 0.85 μm optical spot size. At the same time, individual pixels retain a dense, multi-peak structure, implying an unresolved local spectral manifold below optical resolution. Correlations among centroid, dominant energy, asymmetry, width, entropy, sharp fraction, and roughness indicate that the micron-scale energy landscape and local manifold complexity can be statistically separated, while remaining correlated across the map, consistent with a hierarchical organization of the emission spectrum. These results establish hierarchical inhomogeneity as an organizing principle of $MoSe_2/WSe_2$ moiré superlattice photoluminescence.

Moiré superlattices in twisted or lattice-mismatched van der Waals heterostructures have emerged as a highly tunable platform for quantum materials. In moiré superlattices, the long-wavelength moiré potential can reorganize electronic and excitonic states while remaining controllable through twist angle, stacking registry, strain, and electrostatic environment.[1-4] In transition metal dichalcogenide heterobilayers, this tunability has enabled interlayer moiré excitons with modified optical selection rules, localization, and opportunities for correlated excitonic phases.[2,5-8] For example, periodic moiré trapping has motivated lattice-boson descriptions of excitons, including moiré Bose–Hubbard physics, and has enabled the realization of interaction-driven ordered excitonic states, exciton crystals.[9-11] At the same time, the possibility of spatially ordered moiré-trapped excitons has made these systems attractive platforms for quantum-emitter arrays.[12]

Among these systems, interlayer excitons in $MoSe_2/WSe_2$ heterobilayers provide a rich platform for studying moiré-modulated optical responses, localization, and exciton interactions. At low temperatures, however, the photoluminescence (PL) spectrum of this moiré superlattice, particularly at the interlayer-exciton energy region, is often very complex, featuring a broad emission envelope with many sharp peaks.[7,13,14] In $MoSe_2/WSe_2$, a direct microscopic identification of each individual line is challenging because the observed PL can involve multiple interlayer channels and phonon-assisted emission from momentum-indirect reservoirs, while narrow localized emissions can persist even when the intrinsic moiré potential is suppressed, and donor–acceptor-pair excitons have also been suggested as sources of dense, sharp emission spectra.[15,16] Instead of assigning each line, we ask a different question about this system: how is the spectral complexity organized in real space?

Here, we report a hyperspectral PL image measurement[17] combined with peak-decomposition-free statistical analyses to address the nature of spectral complexity. We analyze a cryogenic hyperspectral PL map of a $MoSe_2/WSe_2$ heterobilayer acquired on a 20 × 20 grid with a 400 nm spatial pitch. From each spectrum, we extract nine peak-decomposition-free descriptors: logarithmic integrated intensity $I_{\mathrm{tot}}$, centroid energy $E_{\mathrm{cent}}$, dominant energy $E_{\mathrm{dom}}$, centroid–dominant offset $\Delta E_{\mathrm{cd}}$, quantile width $W_{80}$, low/high ratio $R_{\mathrm{HL}}$, sharp fraction $F_{\mathrm{S}}$, roughness $R_1$, and spectral entropy $S_{\mathrm{spec}}$; see Supporting Table 1 and Supporting Note 1 for details of the descriptors and see Supporting Table 2 for details of parameters used. We then combine principal component analysis (PCA), Gaussian mixture clustering, spatial correlation analysis, whole-spectrum cosine similarity, and connected-component statistics to determine whether spectral complexity is governed by a single disorder scale or by a hierarchy of spatial scales. Before PCA, all clustering descriptors were standardized to zero mean and unit variance. The Gaussian-mixture clustering was then performed in the PCA-reduced representation.

The $MoSe_2/WSe_2$ heterobilayer was assembled using a standard polymer-based dry-transfer method with flakes of predetermined crystal orientations, determined from the polarization-angle dependence of the second-harmonic generation intensity; we used hexagonal boron nitride flakes to encapsulate the $MoSe_2/WSe_2$ moiré heterobilayer to suppress spatial inhomogeneity. The $MoSe_2/WSe_2$ heterobilayer had an estimated twist angle of $\theta = 60° \pm 3.0°$, determined from post-assembly polarization-resolved second harmonic generation (SHG); SHG showed nearly coincident armchair-axis angles between the monolayer and heterobilayer regions and a reduced heterobilayer SHG intensity, consistent with destructive interference near H-type stacking. Figures 1(a)–1(c) show spatial maps of $A_{tot}$, $E_{cent}$, and $S_{spec}$ across the heterobilayer. Because the hyperspectral map includes regions outside the sample, we excluded pixels with low integrated PL intensity (< 35 counts), leaving 340 spectra for analysis. As clearly seen in Figs. 1(a)-1(c), there are visible micrometer-scale spatial variations in all maps. The optical spot employed has a full width at half maximum of 0.85 μm (700 nm excitation focused with a 0.42-NA objective lens), which is larger than the 400 nm mapping pitch, indicating that neighboring pixels are not independent measurements. The spatial maps should therefore be interpreted as PSF-convolved descriptors of the spectral landscape observable at the present optical resolution. Figure S1 shows six other maps demonstrating similar length-scale spatial variation. Representative single-pixel spectra shown in Fig. 1(d) exhibit sharp peaks in the energy range of 1.25–1.40 eV with relatively broad envelopes, which vary markedly from position to position.[13,15,16]

To visualize the dominant spectral motifs in a compact form, we use a $K = 3$ Gaussian-mixture representation in the PCA-reduced descriptor space. The first two principal components account for 58% of the total variance (PC1: 31.2%, PC2: 26.5%, see Fig. S5(a) for details). Model-selection analysis over different $K$ values does not identify $K = 3$ as a uniquely optimal partition, indicating that the spectral distribution contains additional internal structure (Fig. S2). We therefore regard the $K = 3$ result not as a determination of three physical phases, but as a coarse-grained representation of the major spectral motifs. Direct comparison of the real-space maps for $K = 2$–$5$ shows that changing $K$ primarily changes the degree of subdivision, while the spatially organized mesoscopic pattern remains (Fig. S3). Thus, the central conclusion does not depend on the precise choice of $K$. In this coarse-grained representation, the three families occupy distinct regions in the PCA-reduced descriptor space (Fig. 2(a)) and form spatially contiguous regions in real space (Fig. 2(b)). The family-averaged spectra (Fig. 2(c)) and descriptor means (Fig. 2(d)) show that these groups capture different spectral motifs. We emphasize that the family labels are descriptive data-driven groups and do not imply three distinct thermodynamic phases. Threshold-sensitivity analysis using integrated-PL-area thresholds of 25, 35, and 45 counts shows that the local domain boundaries vary slightly, but the mesoscopic spectral-family organization remains qualitatively unchanged (Fig. S4).

The central quantitative result is that the spatial organization of these spectral families occurs on a micron-scale length scale that exceeds the optical spot size, as indicated in Figs. 1(a)–(c) and Fig. 2(b). The feature-wise correlation decays in Fig. 3(a), the whole-spectrum similarity in Fig. 3(b), the extracted characteristic distances in Fig. 3(c), and the family-resolved domain sizes in Fig. 3(d) all point to the same conclusion; definitions of each quantity are described in the supplementary material. For descriptors showing a positive, approximately monotonic short-range decay of the correlation proxy, feature-wise correlation analysis yields 1/e correlation lengths of 1–2 μm, for example, 2.00 μm, 1.27 μm, 2.02 μm, and 2.05 μm for $E_{\mathrm{cent}}$, $E_{\mathrm{dom}}$, $F_{\mathrm{S}}$, $R_1$, respectively. Whole-spectrum cosine similarity gives a characteristic decay length of 1.87 μm, consistent with the correlation lengths of descriptors. Connected-component analysis[18] of the family map gives a mean equivalent domain diameter of 1.79 μm and a maximum domain size of 4.90 μm. These values all exceed the 0.85 μm excitation-spot FWHM, indicating that the dominant spatial organization is not set solely by optical blurring. The data are therefore consistent with intrinsic spectral inhomogeneity on near-micron-to-micron length scales.

At the same time, the representative single-pixel spectra in Fig. 1(d) show that each pixel remains highly structured. Each valid pixel shows a dense multi-peak spectrum rather than a single smooth emission band. Analysis of the pooled energies of the detected local maxima shows that several energy bins recur across the map, rather than a single ubiquitous line (Fig. S5(b)). The most frequent recurrent bin, near 1.319 eV, still accounts for only a limited normalized peak count when divided by the number of used spectra (Fig. S5(c)). Likewise, the global histogram of pairwise spacings between detected local maxima shows fine structure superimposed on a broad small-spacing background, but no single spacing emerges as a uniquely dominant preferred value (Fig. S5(d)). Therefore, the data do not support a universal ladder of narrow lines across the entire sample. Instead, they suggest an unresolved local spectral manifold with detailed line content that varies across the sample. Here, we use the term "local spectral manifold" to denote the unresolved, line-rich sub-spot spectral structure contained within a single measured pixel, whose detailed line content varies across the sample. Because this manifold is already visible in spectra averaged over a 0.85 μm spot, its underlying degrees of freedom must exist below the optical resolution or otherwise remain unresolved within a single pixel.

Feature correlations, summarized in the partial-correlation matrix in Fig. 4(a), suggest that the local spectral manifold cannot be reduced to a simple intensity modulation of the micron-scale energy landscape. The strongest pairwise monotonic relationships are visualized in Fig. S6. The strongest correlation in the dataset is between $\Delta E_{cd}$ and $R_{HL}$ around the dominant peak (Spearman $\rho = -0.978$), consistent with changes in the asymmetry of the spectral weight distribution rather than a purely rigid shift of an isolated line. The sharp fraction and roughness are also strongly correlated ($\rho = 0.901$), indicating that both descriptors respond to related aspects of line-rich fine structure. In addition, $W_{80}$ is strongly correlated with spectral entropy ($\rho = 0.846$), suggesting that broader spectra in this dataset are associated with more distributed spectral weight and greater local manifold complexity. Finally, $E_{cent}$ and $E_{dom}$ are correlated ($\rho = 0.788$) but not identical, consistent with a micron-scale energy landscape that shifts the overall spectral envelope while allowing the strongest local feature to vary more locally.

Because several descriptors are mutually correlated, the PCA loadings in Fig. 4(b) should not be interpreted as a ranking of independent physical importance. We therefore use Fig. 4(b) only as a compact visualization of coupled descriptor axes. As a robustness check, we repeated the PCA after removing one descriptor from each selected strongly correlated descriptor pair. Although individual loading signs and PC-axis orientations can change because PC1 and PC2 have comparable explained variances, the two-dimensional score-space geometry was largely preserved (Supporting Table 3). The main physical conclusion is therefore based not on the detailed PCA loadings, but on the spatial correlation and whole-spectrum similarity analyses in Fig. 3, which indicate the coexistence of a mesoscopic spectral landscape and an unresolved line-rich local spectral manifold.

Within the present analysis-defined broad/sharp decomposition, the broad and sharp components do not behave as mutually exclusive channels. Their absolute areas are positively correlated across the map, indicating that line-rich spectra are not formed by replacing a broad background with an independent sharp-emission phase. Instead, the sharp manifold appears to ride atop and co-vary with the same micron-scale emission landscape. This disfavors a simple two-phase competition picture and instead supports a hierarchical picture in which local spectral granularity is embedded within a larger-scale background potential.

In summary, we show that the low-temperature PL of $MoSe_2/WSe_2$ cannot be described by a single disorder scale. Instead, we identify a hierarchy of inhomogeneity: a micron-scale spectral landscape observable at the present optical resolution with characteristic length scales of approximately 1–2 μm, and an unresolved local spectral manifold that produces dense multipeak structure within individual pixels. The longer-scale landscape may arise from slowly varying twist, strain, electrostatic, or reconstruction-related fields, while the local manifold may reflect moiré-potential fluctuations, disorder-localized states, donor–acceptor-like traps, or a combination thereof. Static PL mapping alone does not uniquely determine the microscopic origin of the local manifold, but it clearly rules out a one-scale description of the emission landscape. The main conclusion does not depend on the precise number of spectral families used for visualization; it is supported independently by feature-wise spatial correlations and whole-spectrum similarity. Hyperspectral mapping combined with peak-decomposition-free statistical descriptors, therefore, provides a compact route to identifying the organizing scales of interlayer-exciton emission in moiré heterostructures.

## SUPPLEMENTARY MATERIAL

See the supplementary material for additional information.

R.K. was supported by JSPS KAKENHI Grant No. JP25K22210, JP24H02218, and JP23H05469. D.K. was supported by JSPS KAKENHI Grant No. JP24H01210; Canon Foundation; Murata Science and Education Foundation; Shorai Foundation for Science and Technology.

## AUTHOR DECLARATIONS

### Conflict of Interest

The authors have no conflicts to disclose.

### Author contributions

Nurul Fariha Ahmad: Data curation (supporting); Investigation (lead). Yuto Urano: Data curation (equal); Methodology (equal). Kenji Watanabe: Resources (equal); Writing – review & editing (supporting). Takashi Taniguchi: Resources (equal); Writing – review & editing (supporting). Daichi Kozawa: Writing – review & editing (supporting). Ryo Kitaura: Conceptualization (lead); Supervision (lead); Data curation (equal); Methodology (equal); Writing – original draft (lead); Writing – review & editing (lead).

### Data availability

The data that support the findings of this study are available from the corresponding author upon reasonable request.

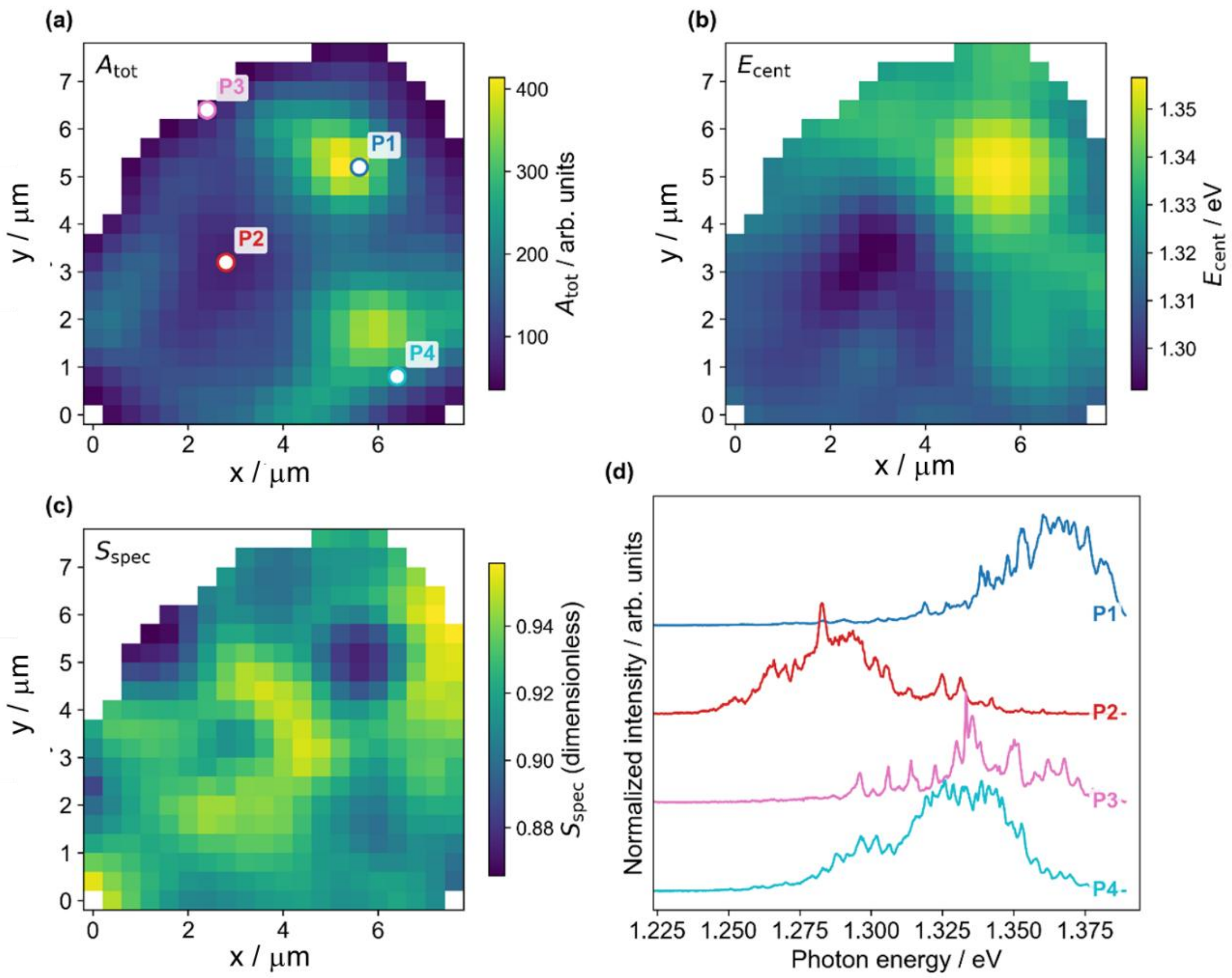

Figure 1. (a) $A_{\mathrm{tot}}$ map of the measured 20 × 20 area. P1–P4 mark four spatially separated pixels selected for spectral display. (b) $E_{\mathrm{cent}}$ map. (c) $S_{\mathrm{spec}}$ map. (d) Normalized single-pixel spectra measured at P1–P4, vertically offset for clarity. The marked pixels exhibit pronounced spectral diversity across the map, while each individual spectrum retains a dense multipeak structure, consistent with an unresolved local spectral manifold at the sub-spot scale. All measurements were performed at 3.5 K.

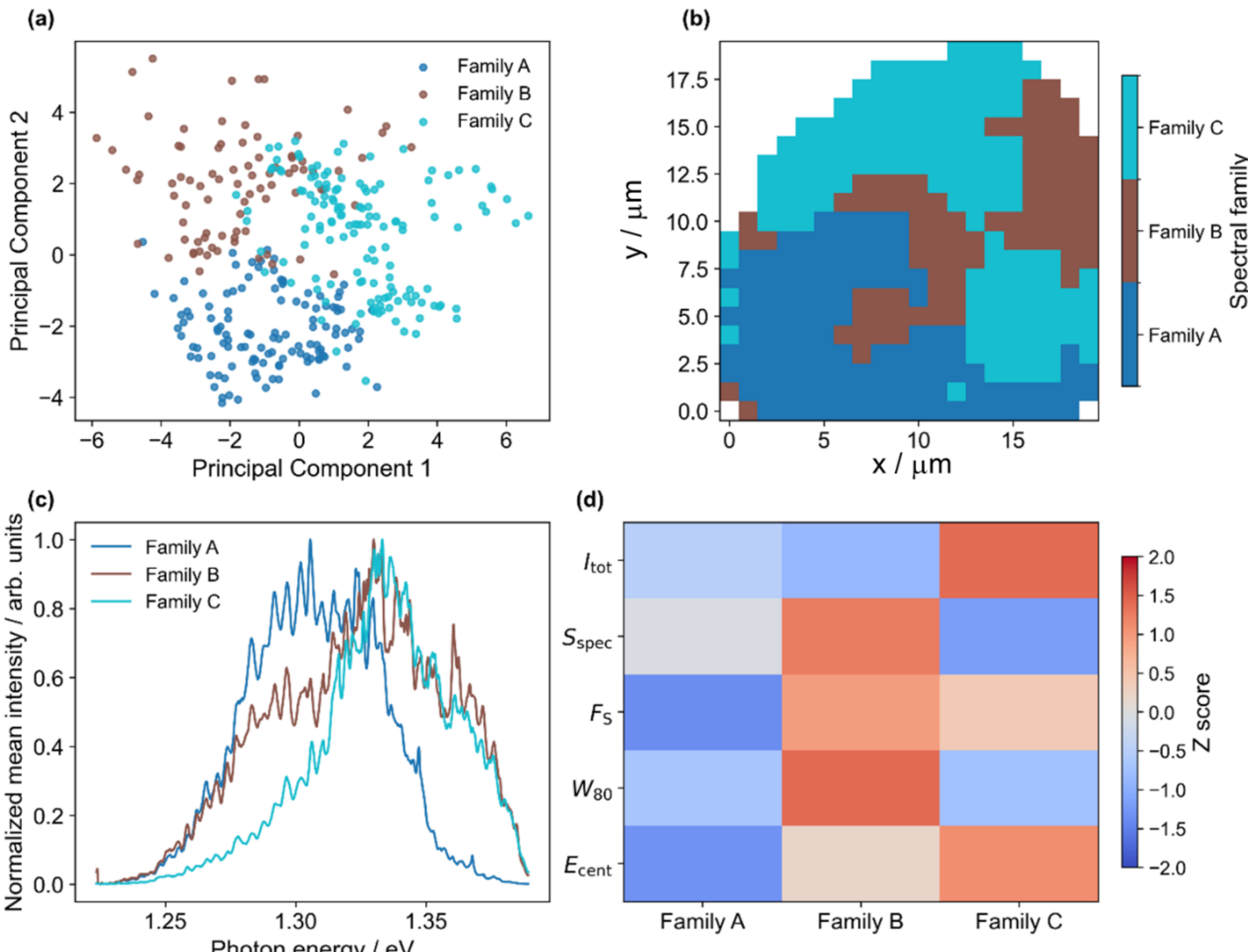

Figure 2. (a) PCA projection of all analyzed pixels, colored by the assigned spectral family. (b) Real-space map of the three families. (c) Family-averaged PL spectra, normalized to emphasize differences in spectral shape. (d) Z-scored family means of selected descriptors, shown using the compact labels $E_{cent}$, $W_{80}$, $F_S$, $S_{spec}$, and $I_{tot}$. The consistency between descriptor-space clustering, the real-space family map, and the family-averaged spectra highlights a compact coarse-grained representation of dominant spectral motifs. These family labels are used descriptively and are not intended to imply three distinct thermodynamic phases.

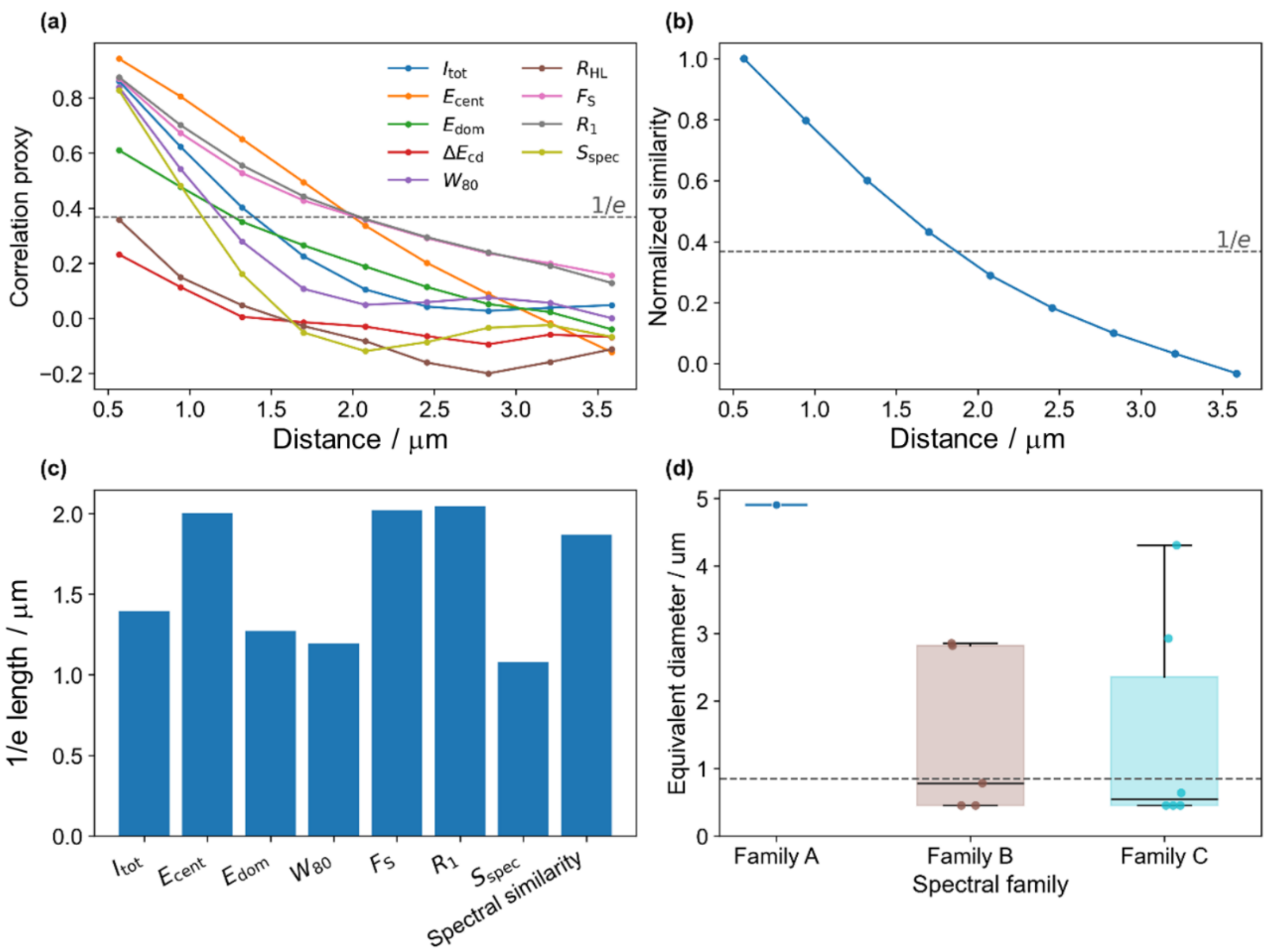

Figure 3. (a) Feature-wise correlation-proxy curves as a function of real-space distance for the nine core descriptors: $I_{\mathrm{tot}}$, $E_{\mathrm{cent}}$, $E_{\mathrm{dom}}$, $\Delta E_{\mathrm{cd}}$, $W_{80}$, $R_{\mathrm{HL}}$, $F_{\mathrm{S}}$, $R_1$, and $S_{\mathrm{spec}}$. (b) Distance dependence of the whole-spectrum similarity, evaluated from the similarity between full single-pixel spectra. (c) Summary of the extracted characteristic $1/e$ distances from the feature-wise correlation-proxy curves and the whole-spectrum similarity curve. (d) Family-resolved equivalent domain diameters obtained from connected components in the spectral-family map. Taken together, these analyses show that the dominant spatial structure of the PL response extends beyond the optical spot size, indicating a micron-scale spectral landscape superimposed on an unresolved sub-spot local spectral manifold. The dotted horizontal line in panels (a) and (b) marks the $1/e$ threshold used to extract the characteristic decay lengths, whereas the dotted line in panel (d) represents the spatial resolution of the current mapping.

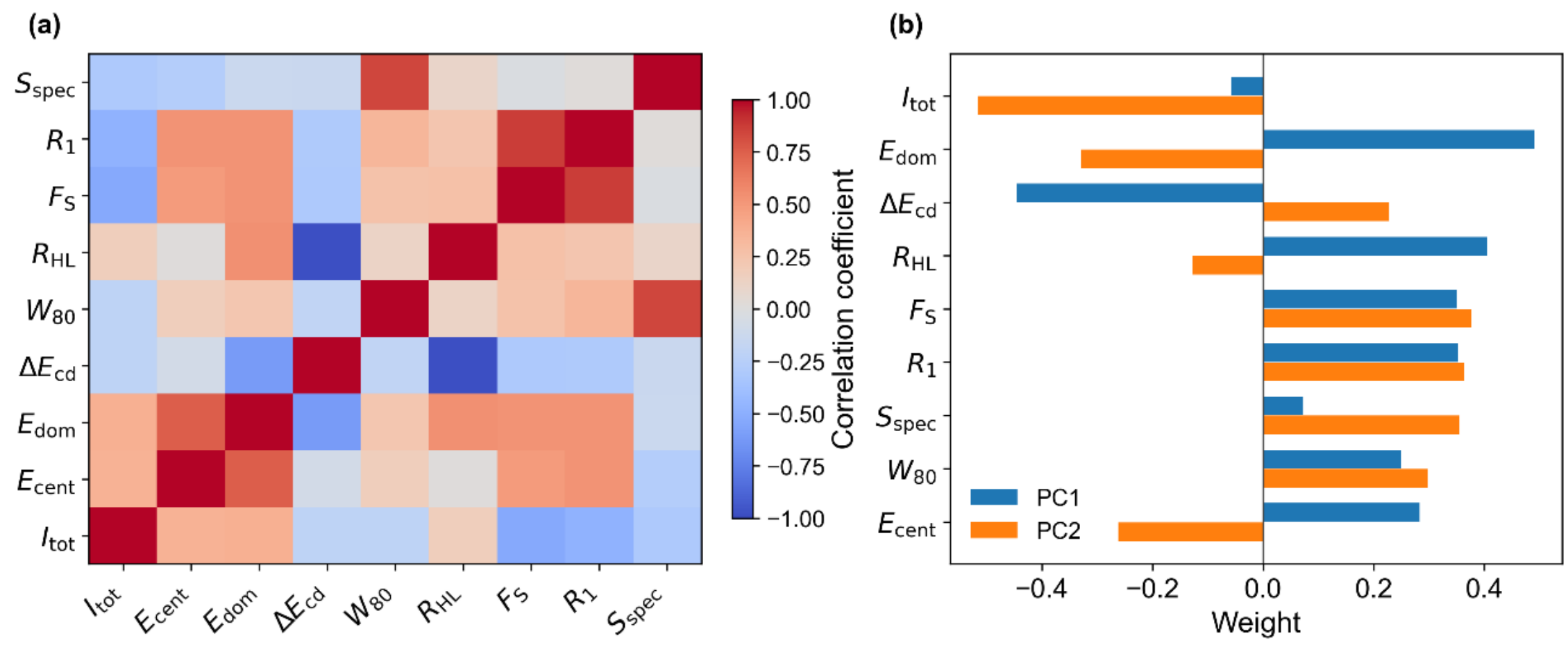

Figure 4. (a) Partial-correlation matrix of the core descriptors, evaluated after controlling for the overall PL intensity. (b) PCA loadings of the same core descriptors. Because several descriptors are mutually correlated, the PCA loadings are used only as a compact visualization of coupled descriptor axes and not as a ranking of independent physical importance.

# Supporting Information

## Hierarchical spectral inhomogeneity in photoluminescence of a twisted $MoSe_2/WSe_2$ heterobilayer moiré superlattice revealed by hyperspectral mapping

*Nurul Fariha Ahmad[1], Yuto Urano[1], Kenji Watanabe[3], Takashi Taniguchi[1], Daichi Kozawa[1], and Ryo Kitaura[1,2,*]*

*[1] Research Center for Materials Nanoarchitectonics, National Institute for Materials Science, 1-1 Namiki, Tsukuba-city, Ibaraki 305-0044, Japan*

*[2] Graduate School of Chemical Sciences and Engineering, Hokkaido University, 5, Kita 8 Nishi, Kita-ku, Sapporo-city, Hokkaido 060-8628, Japan*

*[3] Research Center for Electronic and Optical Materials, National Institute for Materials Science, 1-1 Namiki, Tsukuba-city, Ibaraki 305-0044, Japan*

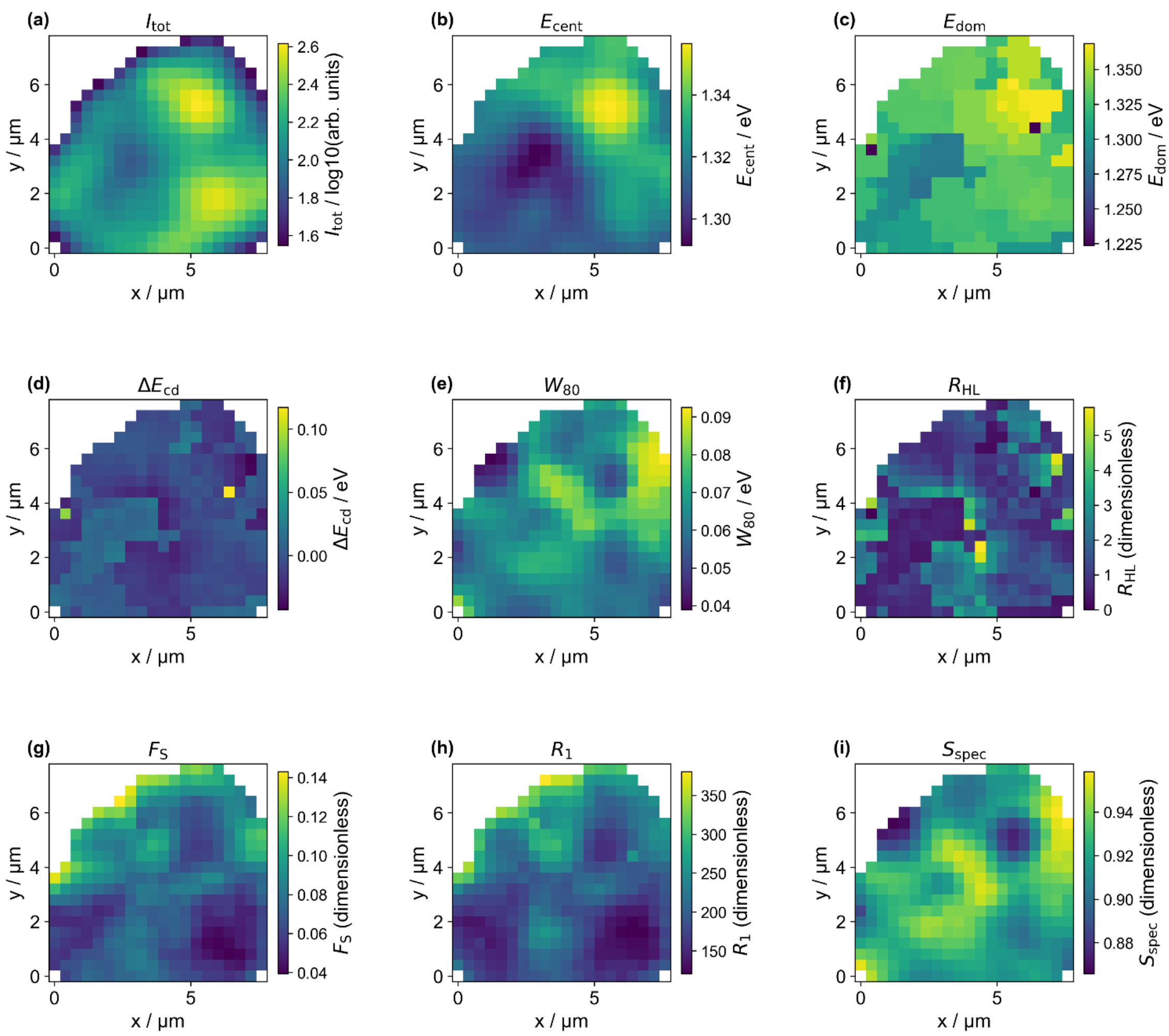

Figure S1. Spatial maps of all peak-decomposition-free descriptors, $I_{tot}$, $E_{cent}$, $E_{dom}$, $\Delta E_{cd}$, $W_{80}$, $R_{HL}$, $F_S$, $R_1$, and $S_{spec}$.

All descriptors show spatial variation on near-micron- to micron-length scales. Pixels outside the analyzed region or below the integrated-intensity threshold are masked. The maps show that both energy-landscape and fine-structure descriptors vary across the heterobilayer on near-micron- to micron-length scales. Descriptor definitions are given in Supporting Table 1.

Alt Text: Spatial maps of nine spectral descriptors extracted from the hyperspectral photoluminescence dataset. Each panel shows a different descriptor over the same measured region, with color indicating the local descriptor value. The maps reveal correlated micron-scale spatial variations across intensity, energy, width, entropy, and fine-structure-related quantities.

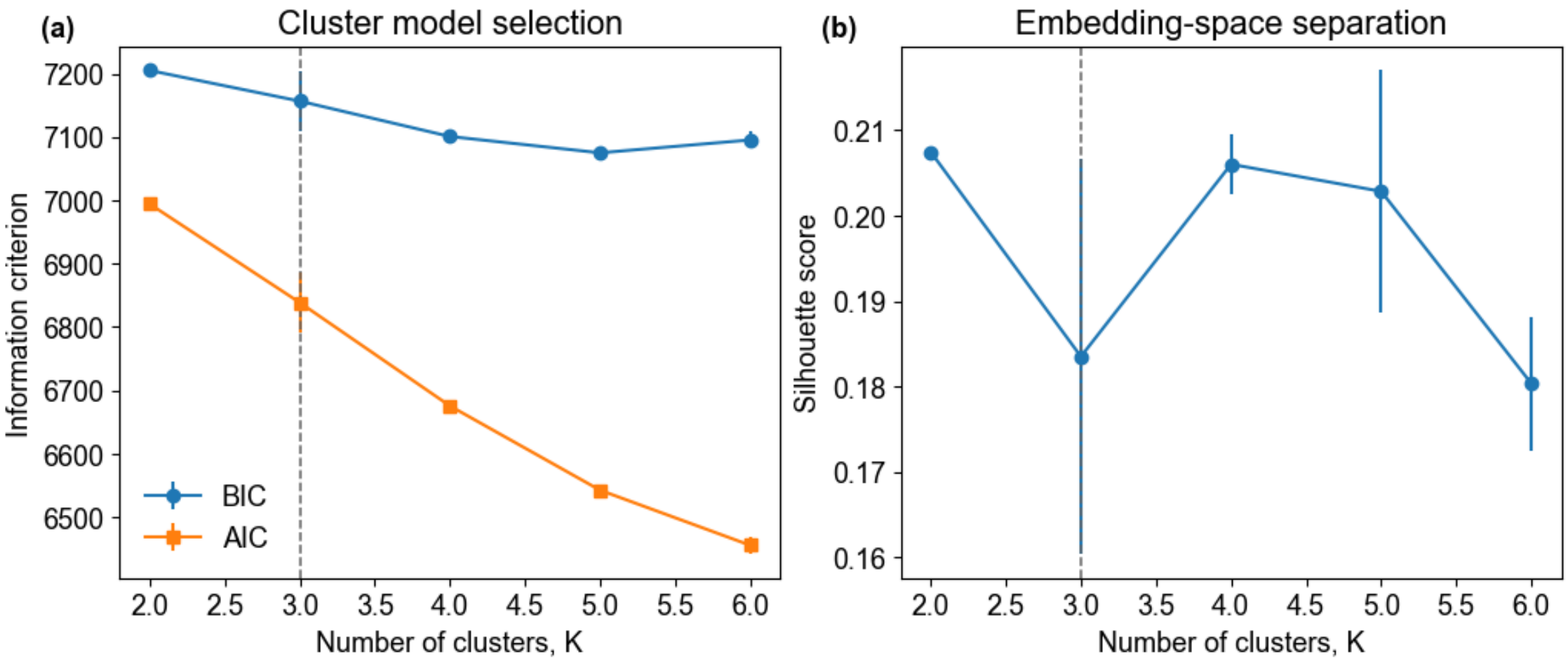

Figure S2. Model-selection analysis for spectral-family decomposition. (a) BIC and AIC for Gaussian-mixture models with different numbers of components K. (b) Silhouette scores in the PCA-reduced descriptor space.

The analysis does not identify K = 3 as a unique optimum. Instead, the continued improvement of the likelihood-based criteria and the modest silhouette scores indicate that the spectral distribution contains additional substructure and is not composed of sharply separated discrete classes. In the main text, K = 3 is therefore used only as a compact coarse-grained representation of the dominant spectral motifs.

Alt Text: Model-selection plots for Gaussian-mixture clustering. The figure compares information criteria and embedding-space separation metrics across different numbers of clusters, showing that K=3 is not uniquely selected as the optimal cluster number.

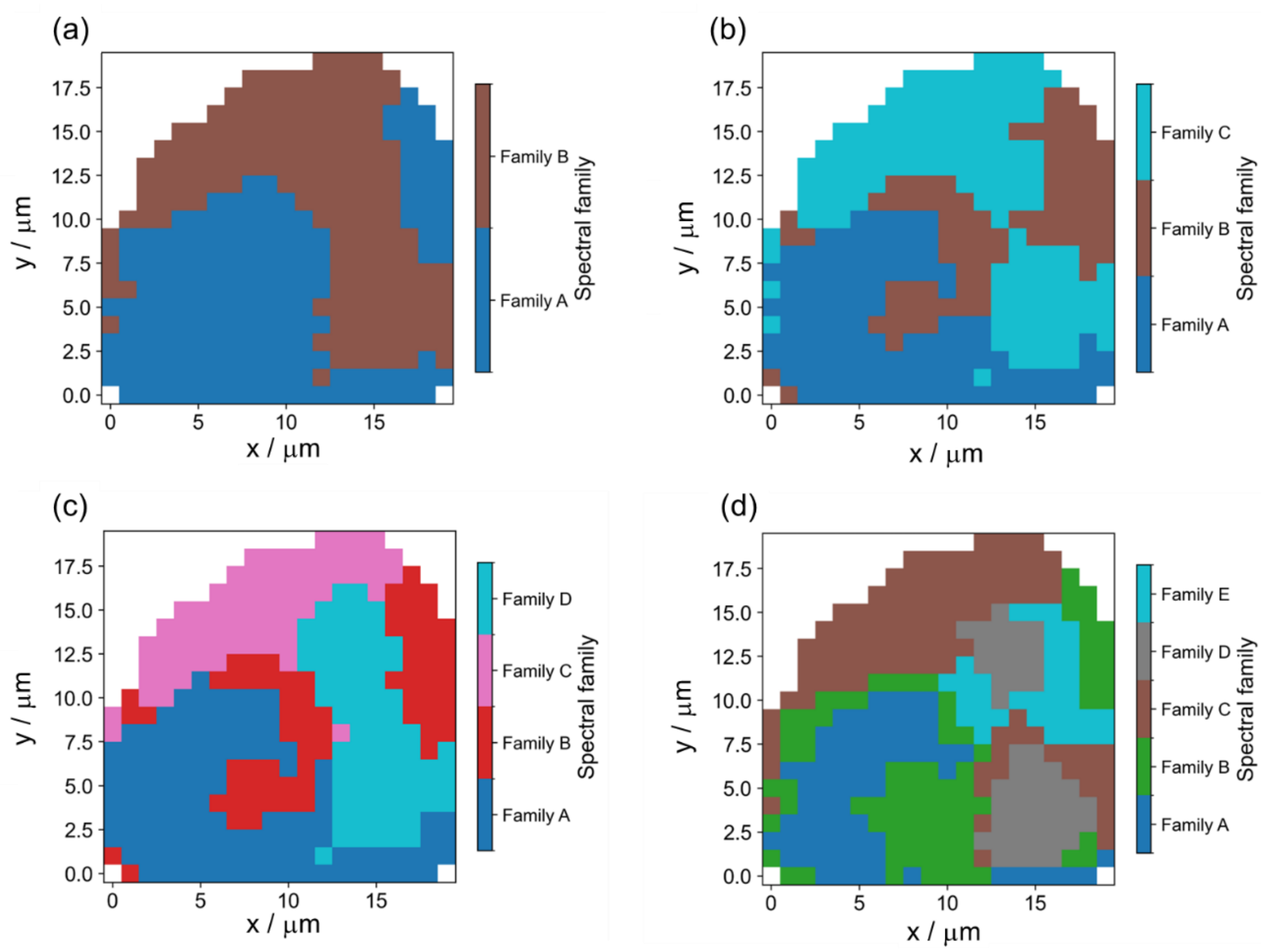

Figure S3. Dependence of the real-space spectral-family map on the number of GMM components.

Real-space maps of the spectral-family labels obtained for $K$ =2, 3, 4, and 5 using the same standardized descriptor set and PCA-reduced embedding. The number of clusters changes the degree of subdivision of the spectral landscape: $K = 2$ yields a coarse partition, whereas $K = 4$ and $K = 5$ further split the same mesoscopic regions. In all cases, the labels form spatially organized domains rather than a random pixel-to-pixel distribution. This confirms that the main conclusion about mesoscopic spectral organization does not rely on the specific choice of $K = 3$.

Alt Text: Real-space spectral-family maps obtained using different numbers of Gaussian-mixture components. The cluster number changes the degree of subdivision, but spatially organized mesoscopic regions remain visible across the tested values.

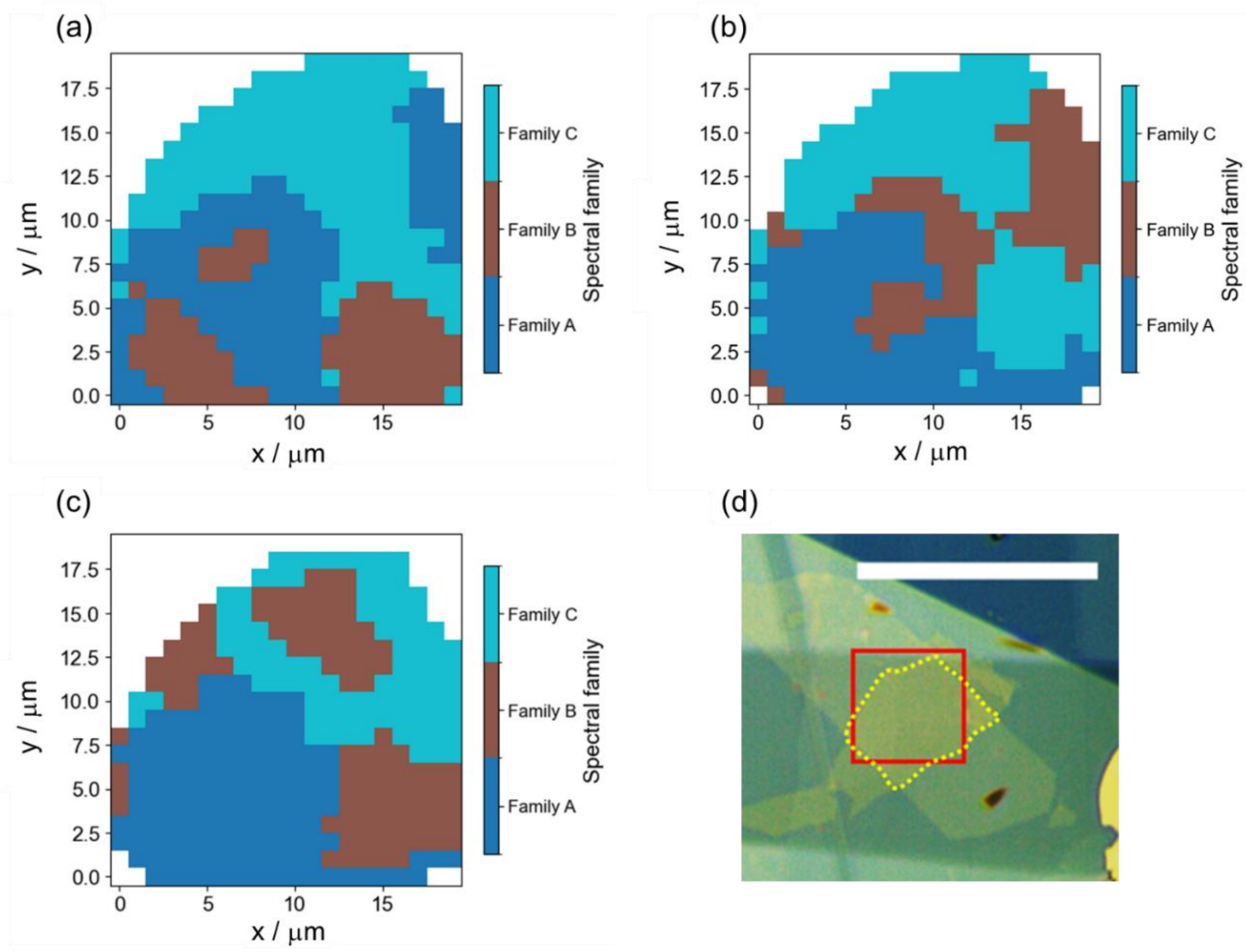

Figure S4. Threshold sensitivity of the spectral-family map.

Real-space spectral-family maps obtained using integrated-intensity thresholds of (a) 25, (b) 35, and (c) 45 to exclude off-sample or near-background pixels. All thresholds show mapping regions consistent with (d) an optical image of the sample. The detailed boundaries and small isolated regions vary with the threshold, but the large-scale mesoscopic organization of the spectral-family map is preserved. This confirms that the main conclusion of spatially organized spectral inhomogeneity is not an artifact of the specific threshold value used in the main analysis.

Alt Text: Threshold-sensitivity analysis of the spectral-family map. The maps compare results obtained using different integrated-photoluminescence-area thresholds. Local boundaries vary slightly, but the large-scale spatial organization is preserved.

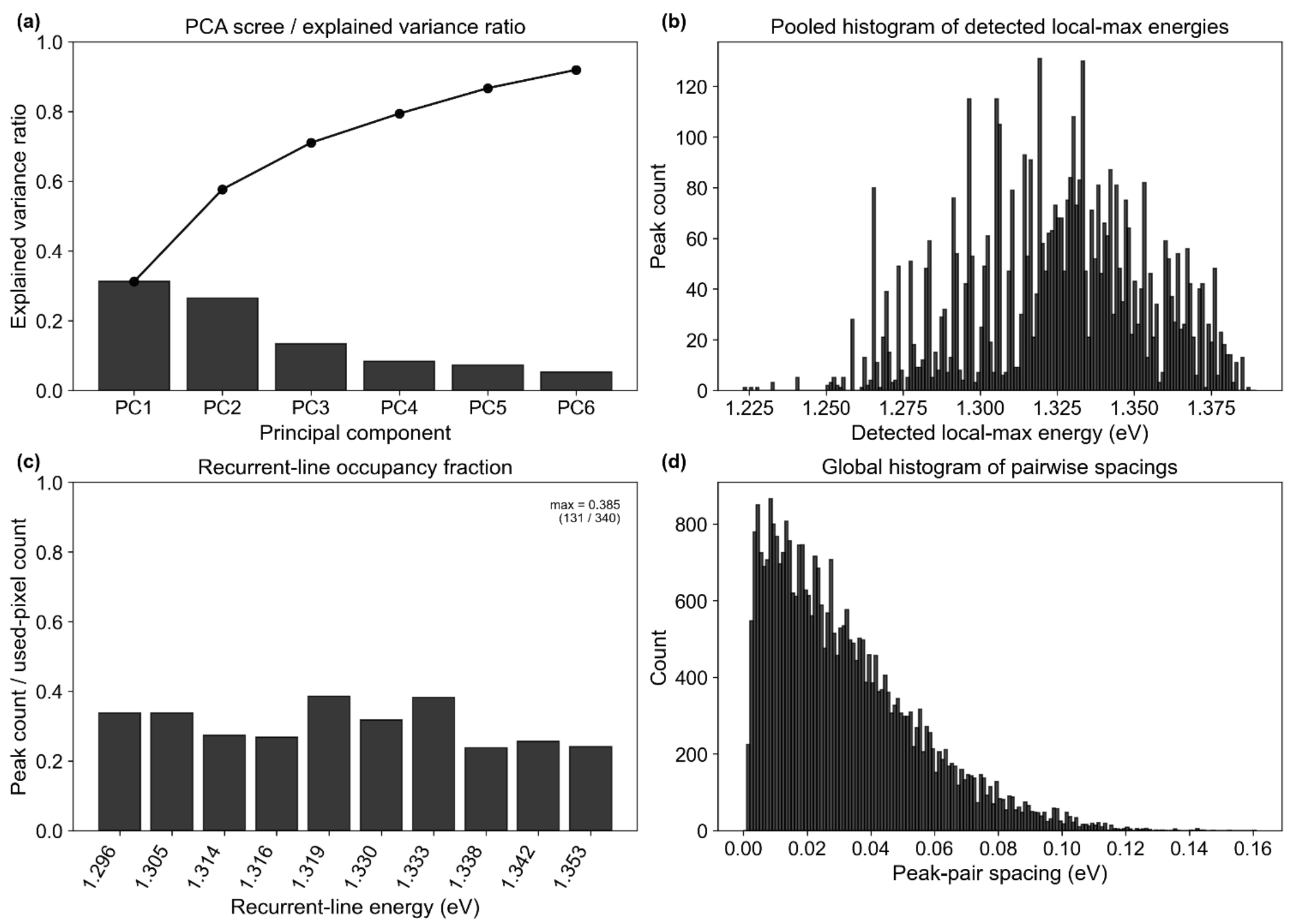

Figure S5. Supporting analyses for the PCA representation and recurrent-line statistics of the hyperspectral photoluminescence dataset. (a) Explained variance ratios of the principal components used in the PCA of the peak-decomposition-free descriptors; PC1 and PC2 account for 31.2% and 26.5% of the total variance, respectively. (b) Histogram of the energies of detected local maxima pooled over all used spectra, showing several recurrent energy bins across the mapped region. (c) Peak-count fractions of the top 15 recurrent energy bins, obtained by dividing the pooled detected-local-max count in each bin by the number of used spectra (340). (d) Histogram of all pairwise spacings between detected local maxima, showing fine structure superimposed on a broad small-spacing background rather than a single sharply dominant preferred spacing.

Alt Text: Supporting analyses of the spectral descriptor space and recurrent spectral features. The panels summarize PCA variance, recurrent-energy features, and global spacing statistics, providing additional evidence for structured spectral complexity beyond isolated peak assignments.

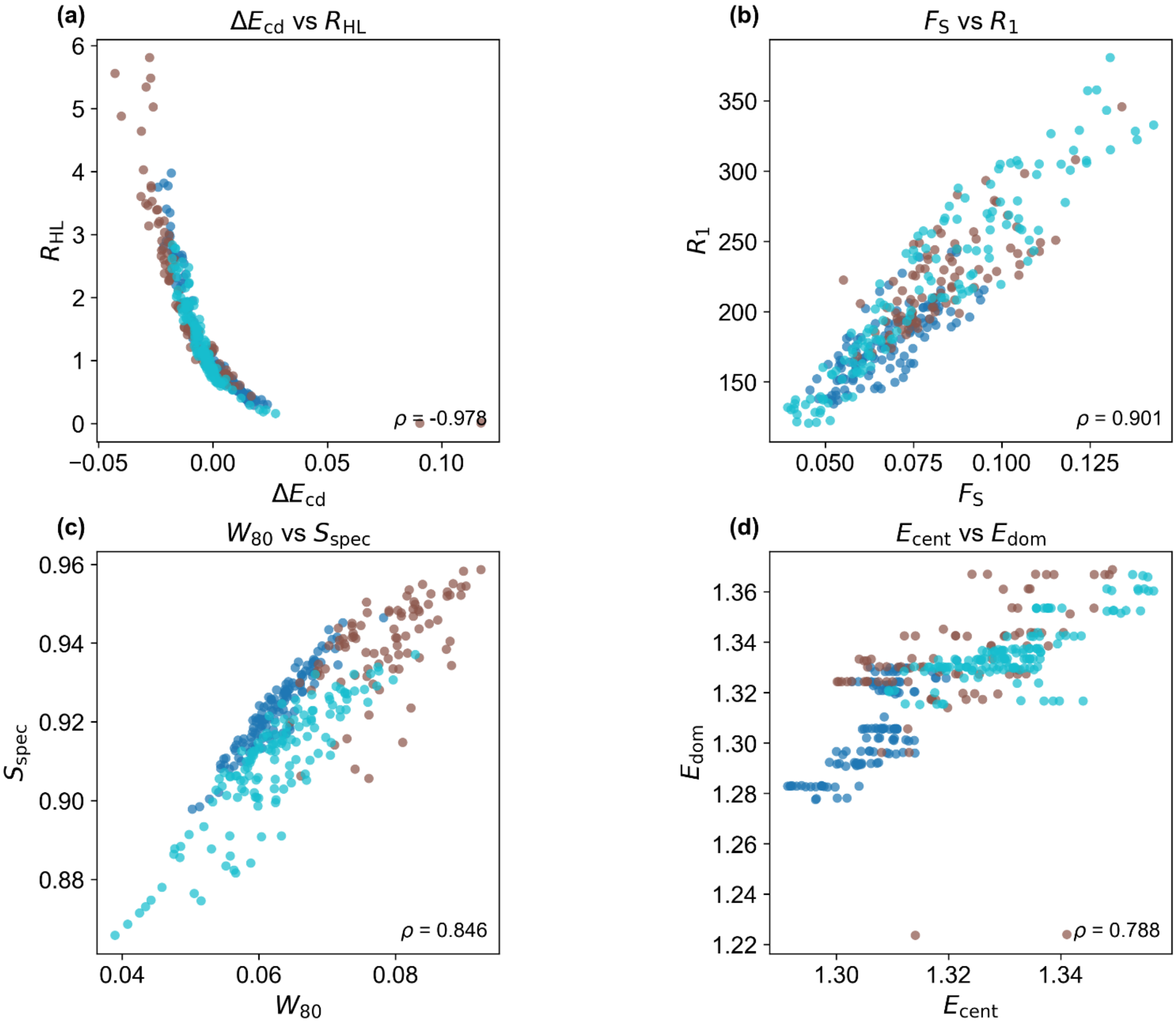

Figure S6. Scatter plots showing the relation between (a) $\Delta E_{cd}$ and $R_{HL}$, (b) $F_S$ and $R_1$, (c) $W_{80}$ and $S_{spec}$, and (d) $E_{cent}$ and $E_{dom}$. Each point corresponds to one used spectrum (340 pixels total). The annotations report the Spearman rank-correlation coefficient $\rho$ for each pair. These plots visualize the monotonic relationships summarized in the correlation analysis used in the main text.

Alt Text: Scatter plots of selected descriptor pairs with strong Spearman correlations. The plots show monotonic relationships among correlated spectral descriptors, illustrating that several descriptors capture related but not identical aspects of spectral variation.

## Supporting Table 1. Core spectral descriptors

| Label | Definition | Interpretation/implementation note | Related refs. |
|---|---|---|---|
| $I_{tot}$ | $\log_{10}(A_{tot})$, $A_{tot}=\int I(E)\,dE$ | Logarithmic integrated PL intensity. Used to compress the dynamic range and as a control variable in partial-correlation analysis. | [S1,S2] |
| $E_{cent}$ | $\int E\, I(E)\, dE \,/\, \int I(E)\, dE$ | First spectral moment (centroid energy); tracks the energy landscape in a smooth, intensity-weighted way. | [S1,S2] |
| $E_{dom}$ | Argmax $E$ $I(E)$ | Energy of the strongest spectral bin. Useful as a simple peak-position summary, but it can jump when the strongest local peak changes identity. | [S1,S2] |
| $\Delta E_{cd}$ | $E_{cent} - E_{dom}$ | Measures asymmetry or unresolved tail weight relative to the dominant peak position. | [S1,S2] |
| $W_{80}$ | $E_{90} - E_{10}$<br>($E_q$: the intensity-weighted q-quantile energy) | Robust width measure based on the central 80% of the spectral weight rather than on a single fitted linewidth. | [S1,S2] |
| $R_{HL}$ | $A(E < E_{dom}) \,/\, A(E \geq E_{dom})$ | Implements the low-energy/high-energy area ratio with the split | [S1,S2] |

| Label | Definition | Interpretation/implementation note | Related refs. |
| --- | --- | --- | --- |
| | | taken at $E_{dom}$; quantifies spectral skewness across the dominant-energy boundary. | |
| $F_S$ | $A_{sharp} / (A_{sharp} + A_{broad})$. | Sharp fraction obtained from a broad/sharp decomposition. The broad component is a nonnegative Savitzky–Golay-smoothed envelope clipped not to exceed $I(E)$, and the sharp component is the positive residual $I(E) − I_{broad}$. | [S7] |
| $R_1$ | $R_1 = (\int \lvert dI/dE \rvert\, dE) / (\int I(E)\, dE)$ | Derivative-based roughness that increases with fine spectral structure and rapid local intensity variation. | [S1,S2,S7] |
| $S_{spec}$ | $−\Sigma_k p_k \ln p_k / \ln N$<br>$(p_k = I_k / \Sigma_j I_j)$ | Normalized Shannon entropy of the discrete spectral-weight distribution; larger values correspond to more distributed spectral weight. | [S2,S3] |

Note: The first six descriptors are standard spectral-summary quantities adapted to a pixel-wise PL spectrum. F_S and R_1 are analysis-specific compactness/fine-structure measures; they are defined explicitly above so that no fitted peak model is required.

Alt Text: Table defining the core spectral descriptors used in the analysis. The table lists each descriptor, its compact notation, mathematical definition, and qualitative physical interpretation.

**Supporting Table 2. Key analysis parameters used in the spectral-statistical analysis.**

Only parameters that directly affect spectral entropy, local maximum detection, recurrent energy analysis, global spacing histograms, and spatial correlation/similarity analyses are listed.

| Analysis item | Key parameter | Value used | Purpose / note |
|---|---|---|---|
| Common spectral window | Photon-energy range and discretization | 1.2236–1.3892 eV; N = 1340; median ΔE = 0.000123 eV | Fixed processed window and binning used for descriptor, entropy, recurrent-energy, and spacing analyses. |
| Preprocessing | Baseline, clipping, smoothing | Baseline = a constant value subtraction using averaged values from 1.225-1.228 eV; negative intensities clipped; Savitzky–Golay pre-smoothing window 5 points, order 2 | Defines the processed non-negative spectra used for descriptor extraction. |
| Local maxima | Detection thresholds | SG window 5 points, order 2; prominence ≥ 0.06×max and 4σ; height ≥ 0.03×max; minimum spacing 0.001 eV | Local maxima are not fitted peaks; actual count: median 14, range 6–24 peaks/spectrum. |
| Recurrent-energy analysis | Global local-max histogram and line assignment | Bin width 0.001 eV; minimum count 4; tolerance ±0.0015 eV; integration half-width ±0.0015 eV; detected lines = 42 | Identifies recurrent local-max energies from pooled spectra without assigning every individual peak. |
| Global-spacing histogram | Pairwise spacing definition and binning | All $\|E_i - E_j\|$ within each valid spectrum; bin width 0.001 eV; | Computed from detected local maxima to test for recurring |

| Analysis item | Key parameter | Value used | Purpose / note |
| --- | --- | --- | --- |
| | | 34,759 spacings; range 0.0010–0.1609 eV | energy separations without line-by-line fitting. |
| Spatial correlation / similarity | Radial lag bins and spatial sampling | 10 lag bins; maximum lag fraction 0.4; pixel pitch 400 × 400 nm; spot FWHM 850 nm | Used for feature-wise correlation proxy and whole-spectrum similarity; results are PSF-convolved spatial statistics. |

Alt Text: Table summarizing key analysis parameters used for spectral preprocessing, descriptor extraction, masking, spatial correlation analysis, recurrent-energy analysis, and global-spacing histogram construction.

**Supporting Table 3. Reduced-descriptor PCA score-space similarity.**

| PCA case | Descriptor removed | No. desc. | PC1+PC2 variance (%) | Procrustes $R^2$ | Linear score-space $R^2$ |
|---|---|---|---|---|---|
| Full 9 descriptors | none | 9 | 59.3 | 1.000 | 1.000 |
| Drop $\Delta E_{cd}$ | centroid–dominant offset | 8 | 60.8 | 0.885 | 0.888 |
| Drop $R_{HL}$ | low/high ratio | 8 | 62.7 | 0.931 | 0.933 |
| Drop $F_S$ | sharp fraction | 8 | 61.2 | 0.846 | 0.854 |
| Drop $R_l$ | roughness | 8 | 61.5 | 0.855 | 0.861 |
| Drop $W_{80}$ | 80% spectral width | 8 | 63.8 | 0.939 | 0.940 |
| Drop $S_{spec}$ | spectral entropy | 8 | 64.6 | 0.938 | 0.940 |

Note: Direct one-to-one comparison of individual loading signs is not used because PCA axes can change sign or rotate when PC1 and PC2 have comparable explained variances. Procrustes $R^2$ and linear score-space $R^2$ quantify preservation of the two-dimensional score distribution relative to the full 9-descriptor PCA.

Score-space similarity was evaluated after removing one descriptor at a time from selected correlated descriptor pairs. The full 9-descriptor PCA is used as the reference. $R^2_{proc}$ was calculated after centering, Frobenius normalization, and optimal orthogonal Procrustes alignment of the reduced-PC score space to the full 9-descriptor PC score space. $R^2_{lin}$ was calculated by least-squares affine prediction of the full 2D PC scores from the reduced 2D PC scores. High values indicate that the two-dimensional score-space geometry is retained, even though individual PCA loadings may rotate, change sign, or exchange order. Values close to 1 indicate that the two-dimensional PCA score geometry is preserved after alignment.

Alt Text: Table summarizing reduced-descriptor PCA score-space similarity. The table reports how well the two-dimensional PCA score space is preserved after removing individual correlated descriptors.

## Supporting Note 1. Spatial-correlation quantities

### SN1.1 Correlation proxy

For any scalar descriptor *z* defined on pixels i = 1, …, N, we first compute the distance-binned semivariogram

$\gamma_z(r) = 1/2 \langle (z_i - z_j)^2 \rangle$, averaged over all pixel pairs with separation r_ij in the lag bin centered at r.

$C_z(r) = 1 - \gamma_z(r) / \mathrm{Var}(z)$

We refer to C_z(r) as a correlation proxy because it is a semivariogram-normalized correlogram-like quantity rather than a formally fitted covariance model. In the implementation used for the manuscript, $C_z(r)$ is clipped to the interval [−1, 1] for display stability. A negative value does not by itself imply a physically meaningful anticorrelation; in finite maps, it can also arise from noise, descriptor discontinuities, or weak nonstationarity. The semivariogram/correlogram construction follows standard geostatistical practice. [S5,S6]

### SN1.2 Feature-wise correlation decay

The feature-wise correlation decay is the curve $C_z(r)$ plotted against real-space distance *r*. This quantity is used to compare how rapidly different descriptors lose short-range spatial similarity. When the short-range decay is positive and approximately monotonic, we summarize it by the empirical 1/e length $\xi_{1/e}$, defined as the first distance at which $C_z(r)$ crosses 1/e, using linear interpolation between neighboring lag bins. Curves that become negative or strongly nonmonotonic are still informative as exploratory descriptors, but are not over-interpreted as evidence for a single robust correlation length. [S5,S6]

### SN1.3 Whole-spectrum similarity

To measure similarity of the full spectral shape, each processed spectrum $I_i(E)$ is first normalized by spectral area and then by its Euclidean norm:

$$A_i = \Sigma_k I_{ik} \Delta E_k, \qquad u_i = I_i / A_i, \qquad v_i = u_i / ||u_i||_2.$$

The pairwise similarity between pixels i and j is then defined as the cosine similarity

$$s_{ij} = v_i \cdot v_j$$

The mean similarity $\bar{s}(r)$ is obtained by averaging $s_{ij}$ over all pairs in a radial lag bin. To remove the finite-distance baseline, the plateau $s_\infty$ is estimated from the last 20 % of valid lag bins, and the normalized similarity shown in the manuscript is

$$S(r) = [\bar{s}(r) - s_\infty] / [\bar{s}(r_1) - s_\infty],$$

where $r_1$ is the first valid lag bin. The similarity length is the first 1/e crossing of $S(r)$. This construction is closely related to cosine-similarity / spectral-angle ideas widely used in hyperspectral analysis, while remaining agnostic to explicit peak assignment. [S1,S2]

**SN1.4 Spearman correlation coefficient**

The Spearman rank-correlation coefficient $\rho_s$ between variables x and y is the Pearson correlation coefficient computed on their ranks R(x) and R(y):

$$\rho_s = \mathrm{corr}[\, R(x), R(y) \,].$$

It measures monotonic association and is less sensitive to strictly nonlinear rescaling than Pearson correlation. In the present workflow, the heatmap labeled "Spearman" is constructed by ranking each variable first and then computing pairwise correlations on the ranked variables. When a partial-correlation heatmap is requested, the ranked variables are residualized against the specified control variable(s)—typically I_tot—and Pearson correlation is then applied to those residuals. [S4]

**Supporting Note 2. Auxiliary analysis methods**

**SN2.1 Family-resolved domain size determination**

**Input.** A 2D map of the spectral-family label at each analyzed pixel.

**Procedure.** For each family, all pixels assigned to that family are converted into a binary mask. Connected components are then identified using 4-neighbor connectivity (up, down, left, right). For each connected component, the area is $A = N_{pix} \times A_{pix}$, where $N_{pix}$ is the number of pixels in the component, and A_pix is the physical area of one map pixel. The reported domain size is the equivalent circular diameter

$$d_{eq} = 2 \sqrt{(A / \pi)}.$$

**Output.** The component table contains the area, equivalent diameter, and row/column span of each domain; the family-resolved summary contains the number of components and the mean/median/max equivalent diameter for each family. Connected-component labeling and region-based size measurement follow standard digital-image-analysis practice. [S9,S10]

**SN2.2 Recurrent-energy analysis**

**Purpose.** To summarize energy bins that recur across many single-pixel spectra without requiring a global peak-assignment model.

**Procedure.** (i) For each processed spectrum, a smoothed working copy is generated, and local maxima are detected using prominence, height, and minimum-distance criteria. (ii) All detected local-max energies from the selected spectra are pooled into a histogram with a user-defined bin width $\Delta E_{\text{bin}}$. (iii) Peaks in this histogram define recurrent energy bins. (iv) For Fig. S5(c), the plotted quantity for each selected bin is the peak-count fraction, defined as the pooled detected-local-max count in that bin divided by the number of used spectra ($N_{\text{used}}$ = 340). The top 15 recurrent bins are displayed.

**Interpretation.** This is an empirical histogram-based recurrence analysis defined for the present work. The cited references are therefore for the generic building blocks—Savitzky–Golay smoothing, local-maximum detection, and histogram-based signal summarization—rather than for an identical pre-existing protocol. [S7,S8,S11]

**SN2.3 Global-spacing histogram**

**Purpose.** To summarize whether the set of detected local maxima exhibits preferred energy separations.

**Procedure.** Within each spectrum, all pairwise positive differences $\Delta E = E_j - E_i$ between detected local-max energies are computed ($j > i$). These pair spacings are pooled over the selected spectra and binned into a global histogram. Peaks or shoulders in the histogram indicate recurrent spacing scales, whereas a broad featureless histogram indicates no strongly preferred spacing. As in the recurrent-energy analysis above, this spacing histogram is an empirical summary defined here; the cited references pertain to peak detection and histogram construction rather than to a canonical moiré-specific protocol. [S8,S11]